\documentclass[12pt]{article}

\usepackage [dvips] {graphicx} 
\newcommand{\be}{\begin{equation}}
\newcommand{\ee}{\end{equation}}
\newcommand{\ba}{\begin{array}}
\newcommand{\ea}{\end{array}}
\newcommand{\bc}{\begin{center}}
\newcommand{\ec}{\end{center}}
\newcommand{\disregard}[1]{{}}
\newcommand{\ti}{\tilde}

\newcommand{\la}{\lambda}

\newcommand{\nt}{\ti{n}}
\newcommand{\mt}{\ti{m}}

\newcommand{\ds}{\displaystyle}

\newcommand{\demi}{{\ds 1\over\ds 2}}

\newcommand{\eps}{\epsilon}
\newcommand{\Nt}{\tilde{N}}
\newcommand{\Mt}{\tilde{M}}

\begin{document}

\title{Generalized stability conditions for binary bose mixtures}
\author{M. Benarous, A. Hocine and A. Mehedi  \\
{\it Laboratory for Theoretical Physics and Material Physics} \\
{\it Department of Physics} \\
{\it Faculty of Exact Sciences and Informatics} \\
{\it Hassiba Benbouali University of Chlef (Algeria)} }

\date{\today}
\maketitle

\begin{abstract}

Considering both interspecies and intraspecies fluctuations in ultracold bose-bose mixtures, we derive generalized stability 
conditions against collapse and phase separation. Furthermore, by examining the energy density of the system, we show that new LHY-like effects 
may enforce or destroy the stability according to the excitation branch, therefore, permitting or prohibiting the existence of droplet states.

\end{abstract}

PACS: 05.30.Jp, 11.15.Tk, 32.80.Pj

\newpage

\setcounter{equation}{0}

\section{Introduction}
Ultracold bose mixtures are raising a great deal of interest owing to their unique properties of the combination 
between intra and interspecies interactions. Experimentally, one can tune these interactions by means of 
Feschbach resonances leading to a rich ground state phase diagram composed of either mixed or separated phases and 
allowing structure formation\cite{cabrera18, semeghini18, lee18}.  

Binary mixtures are generally studied within the mean field approximation by adopting the famous (coupled) Gross-Pitaevskii equations (CGPE).
The latter are quite successful in describing various equilibrium properties of ultracold gases. However, as soon as the mean field energy 
becomes small enough, they become less accurate as quantum fluctuations show up. Since this regime can readily be reached experimentally by finely 
tuning the interspecies interactions, it becomes crucial to analyze the effects of quantum fluctuations on the stability properties 
of the mixture. In the field of ultracold gases, Lee-Huang-Yang (LHY) corrections\cite{LHY57} are among the widely studied fluctuations. Indeed, 
they have been shown to play a major role in the stabilization of a collapsing mixture\cite{Petrov15}. Most importantly, such a phenomenon is 
accompanied with the existence of previously unknown stable states named droplets\cite{semeghini18, dropletexp}. 

In the simplest beyond-Gross-Pitaevskii models for a single condensate, only one-body local fluctuations, denoted generally by 
$\nt$ and $\mt$, are retained. The most widely known are the Time-Dependent Hartree-Fock-Bogoliubov (TDHFB) approximation 
(\cite{Griffin96}-\cite{HUTCH}), the (G)eneralized HFB (taking into account both 
$\nt$ and $\mt$)(\cite{Dalfovo99}-\cite{BB11}), the HFB-Popov (retaining only $\nt$)(\cite{buljan05, merh06, esry97}-\cite{boudjemaa18}), 
and the second order Beliaev approximation\cite{BL58} which introduces $\mt$ in a quite consistent way.
More involved techniques, such as the Bethe Ansatz method, associated with the functional integral 
method\cite{bogol04} and the stochastic projected Gross-Pitaevskii equation\cite{bidasyuk18} have also been successfully used. For a relatively 
small number of particles, the exact many-body problem has also been solved numerically using the Diffusion Monte-Carlo Method\cite{cikojevi18}.
The extended Gross-Pitaevskii equation was also shown in \cite{BPfau19} to yield predictions that differ from the Diffusion Monte-Carlo Method 
indicating that correlations in the system are significant.

Moreover, for multi-species mixtures, going beyond CGPE along the same lines as the methods cited above, requires taking into 
account correlations between atoms of the same species, but also between atoms of different species. In this context, although 
one (and some many-body) correlation functions between atoms or molecules of the same species have been 
extensively studied both experimentally (\cite{hodgman11}-\cite{sch19}) and theoretically (\cite{holland01}-\cite{ng19}), little has been done for 
interspecies correlations in mixtures\cite{WB03, OG20}. It is indeed only recently that they were recognized to play important roles in 
the characterization of quantum phase transitions in mixtures thus relating them to real 
many-body effects (see \cite{penna18} for an illuminating discussion.) For instance, Wang et al.\cite{wang16} studied bose-bose mixtures in 
optical lattices using a perturbative approach, and showed first that interspecies coupling is a necessary condition for interspecies entanglement. 
Second, by computing the entanglement entropy (von Neumann entropy of the reduced density operator), they deduced that it is a clear measure 
of interspecies entanglement. The latter plays an important role in the characterization of the quantum-phase transitions of mixtures, 
specifically in the Mott Insulator to superfluid transition considered there. The entanglement entropy for a two-species Bose-Hubbard dimer 
in a double well was also computed in \cite{lingua18} with the intuitive results that a non-zero hopping causes a non-zero entanglement between 
spatial modes and that the intraspecies interaction contributes to the entanglement. Links et al.\cite{links18} computed the exact ground-state 
correlation functions of an atomic-molecular Bose-Einstein condensate (BEC) and found that the interspecies correlations and variances are 
highly non-trivial in certain regions of the parameter space. Ota et al.\cite{OA20} extended the LHY description in a perturbative way
by computing the ground state energy of a self-bound Bose mixture and Zin et al.\cite{zin221, zin222} elaborated a self consistent description 
of bose-bose droplets. Moreover, Sarkar et al.\cite{Sarkar20} analyzed the interspecies entanglement with impurity atoms in a lattice gas.

On the other hand, focusing on the density matrix, Alon\cite{alon17} found for a harmonic interaction, that the interspecies reduced 
density matrix per particle is separable and given by the product of the intraspecies reduced density matrices per particle. 
The results were compared to \cite{bouvrie14} who found for attractive mixtures vanishing bipartite entanglement between the two species 
and between a single particle (of either kind) and the remaining particles in the mixture. Moreover, Klaiman et al. \cite{kla17} showed in 
the case of an exactly solvable model for a BEC mixture that, while the interspecies two-body density factors to a product of two one-body 
densities in the infinite-particle limit, and the many-body energy and densities coincide with their mean field values, a clear departure from 
the mean field result is observed for the variance of many-particle operators, witnessing that many-body correlations exist in the ground state 
of a trapped mixture whatever the number of particles. 

Using a quite different approach, namely the Multi-layer Multi-Configuration Time Dependent Hartree Bogoliubov (ML-MCTDHB)
(\cite{kro13}-\cite{cao17}), Mistakidis et al.\cite{mistakidis18} studied the quench dynamics of a binary BEC of $^{87}$Rb 
atoms prepared in different internal states. The correlation functions, beyond the mean field, were analyzed up to the one 
and two-body level. Owing to the maximum numbers of orbitals to be used, the authors computed the one and two body inter and 
intraspecies correlation functions for up to 50 atoms. A major challenge to avoid finite size effects and explore the 
thermodynamic limit, is to generalize these results to high enough numbers.

The previous results clearly depict the importance of entanglement in binary mixtures. Indeed, except for very special situations 
such as \cite{pfl09}, there is no a priori argument to claim that interspecies correlations are negligible. In order to account for 
these correlations, we propose in the present work, to use a gaussian density operator for a bose-bose mixture that does not simply 
factorize to a product of density operators for the individual species. The interspecies correlations will result from this 
non-factorazibility. In this spirit, the present approach generalizes the correlated gaussian wavefunctions method (\cite{CGW62}-\cite{CGW16}). 
To proceed further, the dynamics is generated by a variational method (\cite{BV}-\cite{BF99}).
We will henceforth consider a density operator for the entire system that is the exponential of the most general quadratic form in 
the bose field operators. To generate interspecies correlations, this form includes products of fields of different species. 

The paper is organized as follows. In section 2, using the previously described density operator, we derive the energy density 
corresponding to a general two-body Hamiltonian for a homogeneous binary bose mixture at zero temperature. By minimizing the grand 
canonical energy density, we derive generalized expressions for the chemical potentials and the stability 
conditions for the mixture. In order to get more quantitative results, we compute in section 3, the intraspecies and interspecies 
quantum fluctuations for a quasi-one dimensional mixture. While we recover for the former the well-known (negative) LHY terms, we 
show that interspecies fluctuations have the same magnitude and thus confirm that their omission is a quite hazardous 
approximation in the general case. We discuss at the end the significance of these results and their implications 
on the stability of quasi-1D droplets. Some perspectives for future works are also given.

\setcounter{equation}{0}
\section{Stability of the mixture}
Consider a homogeneous mixture composed of two Bose-Einstein condensates at zero temperature. For contact interactions, 
the generally adopted two-body second quantized hamiltonian writes
\be\label{eq1}
\ba{rl}
H&=\sum_{i=1}^2\int_r \psi_i^{+} (r)h_i\psi_i (r)
+\demi\sum_{i=1}^2 g_{ii}\int_{r}\,\psi_i^{+} (r)\psi_i^{+} (r)\psi_i (r)\psi_i (r) \\
&+g_{12}\int_{r}\,\psi_1^{+} (r)\psi_2^{+} (r)\psi_2 (r)\psi_1 (r) ,
\ea
\ee
where $\psi_1^+ (r)$ and $\psi_2^+ (r)$ are the creation bose field operators corresponding to 
the two species satisfying the usual commutation rules $[\psi_i (r),\psi_j^{+} (r^{'})]=\delta_{ij}\delta(r-r^{'})$, 
$[\psi_i (r),\psi_j (r^{'})]=[\psi_i^{+}(r),\psi_j^{+} (r^{'})]=0$. The constants $g_{ii}$ and $g_{12}$ are respectively 
the intra- and interspecies couplings, $m_i$ the masses of the particles and $h_i=-(\hbar ^{2}/2m_i)\Delta$.  

The energy density then writes
\be\label{eq2}
\ba{rl}
{\cal E}&=\demi g_{11}n_1^2+\demi g_{22}n_2^2+ g_{12}n_1n_2 \\
&+\demi g_{11} n_{{c}_{1}}^2\left(2(\Nt_{11}+\Mt_{11})+\Nt_{11}^2+\Mt_{11}^2\right)+\big(1\to 2\big)\\
&+g_{12}n_{{c}_{1}}n_{{c}_{2}}\left(2(\Nt_{12}+\Mt_{12})+\Nt_{12}^2+\Mt_{12}^2\right)
\ea
\ee
where $n_1$, $n_2$ are the total densities and $n_{{c}_{i}}$ the condensate densities. For the sake of clarity, 
we have introduced the reduced "densities" $\Nt_{ij}=\nt_{ij}/\sqrt{n_{{c}_{i}}n_{{c}_{j}}}$ and $\Mt_{ij}=\mt_{ij}/\sqrt{n_{{c}_{i}}n_{{c}_{j}}}$,
where $\nt_{ij}=\langle \psi_i^{+}\psi_j\rangle-\langle \psi_i^{+}\rangle\langle\psi_j\rangle$ and $\mt_{ij}=\langle \psi_i\psi_j\rangle-\langle \psi_i\rangle\langle\psi_j\rangle$. Obviously, $\nt_{ii}$ and $\mt_{ii}$ are the noncondensate and anomalous densities respectively and the 
second line in (\ref{eq2}) (without the quadratic terms) refers, as we will show below, to the well-known LHY contributions. 
Most importantly, the new terms $\Nt_{12}$ and $\Mt_{12}$ appearing in the third line of (\ref{eq2}) are related to the entanglement 
between the two species. It is evident that interspecies interactions are a necessary condition for interspecies correlations \cite{wang16, lingua18}.

The obvious fact is that the contributions of $\Nt_{12}$ and $\Mt_{12}$ are formally comparable to $\Nt_{ii}$ and $\Mt_{ii}$. 
However, for unknown reasons, they are generally omitted in the existing literature 
\cite{Petrov15, PetAst16, LB20, TGAM20, SBSS23, EMS23}. Such assumptions are well justified when the mean field energy dominates. 
On the contrary, when the latter happens to be vanishingly small, as would be the case when one enters the droplet region, 
these assumptions completely break down. Indeed, the finding made by Petrov\cite{Petrov15} that a collapsing mixture 
($g_{12}+\sqrt{g_{11}g_{22}}\to 0$) may be stabilized by quantum fluctuations becomes questionable as the new terms may have 
dramatic effects on the stability of the droplets. It is worth mentioning that these terms have been considered in two recent 
papers. In \cite{RPGA20}, they were written explicitly, but never considered for further calculations. On the contrary, 
the authors of \cite{OG20} considered seriously interspecies fluctuations (which they named mixed anomalous densities), 
but were mainly interested in the Popov approximation at finite temperature. Hence, real quantum fluctuations were 
hidden by thermal effects.


One of the objectives of the present work is to show that the contributions of quantum interspecies fluctuations are not only 
qualitatively, but also quantitatively comparable to the intraspecies fluctuations, even at zero temperature. To this end, 
we will compute $\Nt_{12}$ and $\Mt_{12}$ using the Bogoliubov approach.

Before proceeding further, we begin by writing the general stability conditions of the mixture. The latter are ensured when the 
grand canonical energy density ${\cal E}-\mu_1 n_1-\mu_2 n_2$ reaches a minimum. This requires that
\be\label{stab1}
{\ds\partial {\cal E}\over\partial n_{{c}_{i}}}=0
\ee
and that the Hessian matrix
$$
A_{ij}={\ds\partial^2 {\cal E}\over\partial n_{{c}_{i}}\partial n_{{c}_{j}}}
$$
be positive definite, which means ${\rm Tr}\,A>0$ and ${\rm Det}\,A>0$. The first condition leads to the following expressions for 
the chemical potentials
\be\label{cp1}
\ba{rl}
\mu_1&=g_{11}n_1+g_{12}n_2+g_{11}n_{{c}_{1}}\left(\Nt_{11}+\Mt_{11}\right)+g_{12}n_{{c}_{2}}\left(\Nt_{12}+\Mt_{12}\right)\\
\mu_2&=g_{22}n_2+g_{12}n_1+g_{22}n_{{c}_{2}}\left(\Nt_{22}+\Mt_{22}\right)+g_{12}n_{{c}_{1}}\left(\Nt_{12}+\Mt_{12}\right)\\
\ea
\ee
which are clearly a generalization, taking into account interspecies fluctuations, of the results obtained from the 
extensions of the stationary Gross-Pitaevskii equation \cite{Griffin96, HUTCH, OG20, MOR, CHER03}. 
The second requirements yield the stability conditions against collapse $G_1+G_2>0$ and against phase 
separation $G_1G_2-G_{12}^2>0$, where the generalized coupling constants are defined as follows:
\be\label{ccg}
\ba{rl}
G_1&=g_{11}-g_{12}  {\ds \Nt_{12}+\Mt_{12}\over\ds 2}{\ds n_{{c}_{2}}\over\ds n_{{c}_{1}}}\\
G_2&=g_{22}-g_{12}  {\ds \Nt_{12}+\Mt_{12}\over\ds 2}{\ds n_{{c}_{1}}\over\ds n_{{c}_{2}}}\\
G_{12}&=g_{12} \left(1+{\ds \Nt_{12}+\Mt_{12}\over\ds 2}\right)
\ea
\ee
Obviously, the previous conditions generalize in a quite natural way the well-known stability conditions $g_{ii}>0$, 
$g_{11}g_{22}-g_{12}^2>0$ for a non-correlated homogeneous bose mixture\cite{min74}. Notice that this generalization 
involves the interspecies fluctuations and not the intraspecies ones, which is a quite new result that may have great 
implications on the stability properties of the mixture. For instance, having repulsive intraspecies forces does not 
prevent in general the mixture from collapse. In parameter space, the stability windows against collapse and phase 
separation are strongly affected by the interspecies fluctuations. It is worthwhile noticing that these effects may be 
quite small when one addresses only the question of stability. However, from the expression (\ref{eq2}) of the energy, 
we see that they may have dramatic effects in situations where the mean field energy happens to vanish. This is at the heart 
of the observation made by Petrov\cite{Petrov15}.

In order to make these remarks more quantitative, we will compute in the next section the fluctuations $\Nt_{ij}$ and 
$\Mt_{ij}$ using the Bogoliubov approximation.


\setcounter{equation}{0}
\section{Single Particle Excitations and Stability of Droplets in Quasi-One Dimension}

The single particle excitations may be analyzed by examining the small amplitude motion around static equilibrium. Decomposing the 
shifted operators $\bar{\psi_i}=\psi_i-\langle\psi_i\rangle$ according to Bogoliubov's prescription, we may write in general
\be\label{eq4}
\ba{rl}
\nt_{ij}&=\phantom{-} \sum_k\left[f_k U_k^{(i)}U_k^{(j)}+(1+f_k)V_k^{(i)}V_k^{(j)}\right]
\\
\mt_{ij}&=-\sum_k\left[(1+2f_k)U_k^{(i)}V_k^{(j)}\right]
\ea
\ee
where $U_k^{(i)}$ and $V_k^{(i)}$ are the quasi-particles amplitudes, and $f_k=(e^{\hbar\omega_k/k_B T}-1)^{-1}$ is the 
Bose-Einstein distribution function for the energy mode $\hbar\omega_k$. Evidently, $f_k=0$ at zero temperature, which 
we shall consider from now on.

Using the variational method cited above\cite{BV, BM91, BF99}, we finally obtain the Hartree-Fock-Bogoliubov-De Gennes (HFB-BdG) 
equations \cite{BM13, BB11, RPGA20} in their generalized form 
\be\label{hfbstatic}
\ba{rl}
\hbar\omega_k U_k^{(i)} &=\phantom{-}e_{k_{i}} U_k^{(i)} -g_{ii}\kappa_{ii} V_k^{(i)}
+g_{12}\left(\eta_{12}U_k^{(3-i)}-\kappa_{12}V_k^{(3-i)}\right)\\
\hbar\omega_k V_k^{(i)} &=-e_{k_{i}} V_k^{(i)}+g_{ii}\kappa_{ii} U_k^{(i)} -g_{12}\left(
\eta_{12}V_k^{(3-i)}-\kappa_{12} U_k^{(3-i)}\right)\\
\ea
\ee
where $e_{k_{i}}=\hbar^2k^2/2m_i+g_{ii}n_{c_{i}}\left[1-\Mt_{ii}-\la_i\frac{n_{c_{3-i}}}{n_{c_{i}}}\left(\Nt_{12}+\Mt_{12}\right)\right]$, $\la_i=g_{12}/g_{ii}$, $\eta_{ij}=\sqrt{n_{{c}_{i}}n_{{c}_{j}}}\\(1+\Nt_{ij})$ and $\kappa_{ij}=\sqrt{n_{{c}_{i}}n_{{c}_{j}}}(1+\Mt_{ij})$, 
$i,j=1,2$. 
For the sake of clarity, we will consider from now on a two-component balanced system, for which $m_1=m_2=m$, $g_{11}=g_{22}=g$ 
and $n_1=n_2=n$. Then, $n_{{c}_{1}}= n_{{c}_{2}}=n_{c}$, $\Nt_{11}=\Nt_{22}=\Nt$, $\Mt_{11}=\Mt_{22}=\Mt$ and $\la_1=\la_2=\la$. 
The Eqs. (\ref{hfbstatic}) yield two branches ($\ti{\omega}=\hbar\omega_k/gn_c$):
\be\label{spectrum}
\ba{rl}
\tilde{\omega}_{-} &=\sqrt{\eps_k\left(\eps_k+2(1-\la)-2\la(\Nt_{12}+\Mt_{12})\right)}\\
\tilde{\omega}_{+} &=\sqrt{\eps_k\left(\eps_k+2(1+\la)\right)}\\
\ea
\ee
with $\eps_k=\frac{\hbar^2k^2}{2mgn_c}$. Notice that $\tilde{\omega}_{+}$ does not depend on the interspecies 
correlations and that the symmetry $\tilde{\omega}_{-} (-\la)=\tilde{\omega}_{+} (\la)$ (valid when $\Nt_{12}$ and $\Mt_{12}$ vanish) 
is broken. The quasi-particle amplitudes read now $U_k^{(2)}=-U_k^{(1)}$ and $V_k^{(2)}=-V_k^{(1)}$ for the ($-$) branch, and
$U_k^{(2)}=+U_k^{(1)}$ and $V_k^{(2)}=+V_k^{(1)}$ for the ($+$) branch. This yields the following remarkable solutions: 
$\Nt_{12}=-\Nt$, $\Mt_{12}=-\Mt$ for $\tilde{\omega}_{-}$ and $\Nt_{12}=+\Nt$, $\Mt_{12}=+\Mt$ for $\tilde{\omega}_{+}$. 
Plugging these solutions into the energy density (\ref{eq2}), one gets  
\be\label{en1}
{\ds {\cal E}_{\mp}\over\ds gn^2}=1+ \la +2(1\mp \la)\left(\Nt+\Mt\right)+(1+ \la) \left(\Nt^2+\Mt^2\right).
\ee
One may get a better insight by comparing (\ref{en1}) to its analog in the absence of interspecies fluctuations (called for brevity uncorrelated):
\be\label{enwithout}
{\ds {\cal E}\over\ds gn^2}=1+ \la +2\left(\Nt+\Mt\right)+ \Nt^2+\Mt^2.
\ee
The key point at this level is that, for $\la\simeq -1$ (droplet regime), the LHY term is doubled in the ($-$) sector whereas 
it is canceled by interspecies fluctuations in the ($+$) sector. Since it is assumed that this (positive in 3D) correction stabilizes 
a collapsing condensate\cite{Petrov15}, we may infer from these considerations that, in the ($+$) sector, it will no longer be the case 
in the presence of interspecies fluctuations. 

In quasi-1D geometries, the situation is different as the LHY term is negative\cite{PetAst16, LB20, TGAM20} and hence, an uncorrelated 
mixture may in principle undergo instability as soon as the mean field energy vanishes. However, when the mean field is sufficiently
close to zero, there will be a possibility for a self bound stable state. 



Let us see this more concretely by computing the various fluctuating terms in 1D (or quasi-1D). Plugging the solutions of (\ref{hfbstatic}) into 
(\ref{eq4}), we get:
\be\label{fluct}
\ba{rl}
\Nt &=\frac{1}{2\pi}\left(1-\frac{{\rm arcsinh}(1)}{\sqrt{2}}\right)\gamma_{1D}\sqrt{1\mp\la} \\
\Mt &=-\frac{\sqrt{2}}{8}\gamma_{1D}\sqrt{1\mp\la} \\
\ea
\ee
where $\gamma_{1D}=\sqrt{mg/\hbar^2 n}$ is the Lieb-Liniger gas parameter. Hence, as mentioned above, the LHY term $\Nt +\Mt\simeq -0.117 \,\gamma_{1D}\sqrt{1\mp\la}$ is effectively negative\cite{PetAst16, TGAM20, SBSS23} and falls in the stability region defined in section 2. 
Moreover, one may easily check that $\tilde{\omega}_{-}$ corresponds to the lowest branch for any $\la >0$ while $\tilde{\omega}_{+}$ 
corresponds to the lowest branch for any $\la <0$. 

When approaching the droplet regime by taking $1+\la=\delta g/g<<1$, the energy densities (\ref{en1}-\ref{enwithout}) write, up to first order 
in the fluctuations, 
\be\label{en+}
\ba{rl}
{\cal E}_{+}&\simeq\delta g n^2 -\gamma n^{3/2}\delta g^{3/2}\\
{\cal E}&\simeq\delta g n^2-\gamma n^{3/2}g\delta g^{1/2}\\
\ea
\ee
where $\gamma =0.234\sqrt{m/\hbar^2}$. Hence, although the two expressions have the same functional form in density characterizing the existence 
of a stable self-bound state\cite{PetAst16, LB20, TGAM20}, they differ by their structure in the coupling constants. Indeed, we can easily show 
that the equilibrium density and the minimum energy for the correlated case are smaller by a factor of $(\delta g/g)^2$ and $(\delta g/g)^4$ 
respectively with respect to their uncorrelated values. Hence, interspecies fluctuations shift the equilibrium density and the minimum energy 
to almost zero, forbiding the existence of a self-bound droplet. This is clearly depicted on Fig.1, where we have set $m=\hbar=1$, $\delta g=0.01$ 
and $g=1$ for the sake of a better visualization.
\begin{figure}
\begin{center}
\includegraphics[scale=0.9]{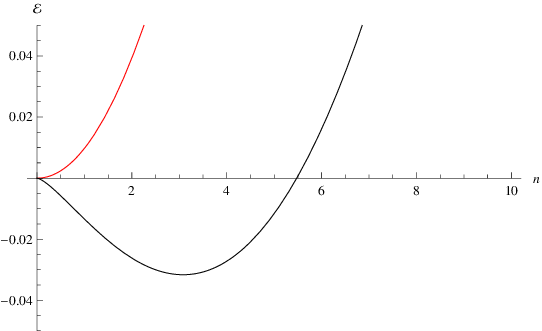}
\caption{Energy (arbitrary units) vs. density in the ($+$) sector (\ref{en+}) for the correlated (${\cal E}_+$: red) and uncorrelated ($\cal E$: black) case. $m=\hbar=1$, $\delta g=0.01$, $g=1$.}
\end{center}
\end{figure}

On the other hand, the ($-$) branch reveals a more interesting behavior, since the energy densities (\ref{en1}-\ref{enwithout}) write in 
this case\be\label{en-}
\ba{rl}
{\cal E}_{-}&\simeq\delta g n^2 -\sqrt{8}\left(1-{\ds 3\over\ds 4}{\ds\delta g\over\ds g}\right)\gamma (gn)^{3/2}\\
{\cal E}&\simeq\delta g n^2-\sqrt{2}\left(1-{\ds 1\over\ds 4}{\ds\delta g\over\ds g}\right)\gamma (gn)^{3/2}\\
\ea
\ee
The minimum of energy writes ${\cal E}_{-}^{(min)}=-1/3\, n_{-}^2\delta g$ where the liquid phase equilibrium density is given by
\be\label{min}
n_{-}\simeq\frac{9\gamma^2}{2} \frac{g^3}{\delta g^2} .
\ee
One easily shows that ${\cal E}_{-}^{(min)}=16\,{\cal E}^{(min)}$ and $n_{-}=4\,n_0$, where $n_0$ is the equilibrium density 
for the uncorrelated case. As can be inferred from Fig.2, this minimum is drastically shifted downward leading to a more 
stable self-bound droplet state in the liquid phase. 

\begin{figure}
\begin{center}
\includegraphics[scale=0.9]{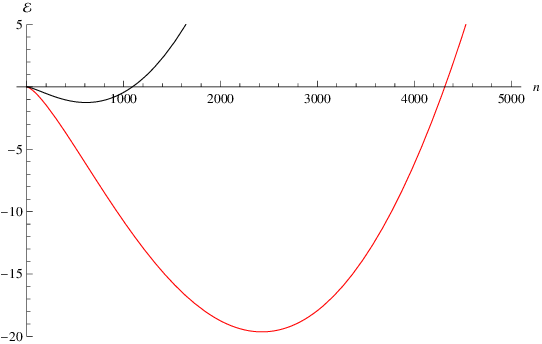}
\caption{Energy (arbitrary units) vs. density in the ($-$) sector (\ref{en-}) for the correlated (${\cal E}_-$: red) and uncorrelated ($\cal E$: black) case. $m=\hbar=1$, $\delta g=0.01$, $g=1$.}
\end{center}
\end{figure}

\setcounter{equation}{0}
\section{Conclusions}

In conclusion, contrary to the abundant scientific literature on binary bose mixtures, where beyond mean field effects consist only  
in intraspecies quantum fluctuations, the so-called LHY corrections, we demonstrate in this work that completely different behaviors 
may show up when one takes into account interspecies fluctuations. In a first step, we derive generalized stability conditions against 
collapse and phase separation taking account of the interspecies fluctuations. These conditions nicely generalize the well known conditions 
in the absence of interspecies fluctuations. Moreover, in the framework of Bogoliubov's approach, the interspecies fluctuations 
are shown to be of the same order of magnitude than the LHY terms. Thus, omitting them, as it is generally assumed, is no longer justified 
in the general case. Most importantly, depending on the excitation branch in the vicinity of the droplet regime, 
the interspecies fluctuations are shown to either enforce the LHY effects or completely cancel them which may lead to the destabilization of 
the mixture. 

In order to better apprehend the impact of these novel behaviors, the previous calculations will be generalized in a forthcoming work to 
the unbalanced non-symmetric case, where one should observe the effects associated with unequal populations and interactions. Furthermore, 
the effects of geometry of the system should also be examined carefully as the LHY corrections are not at all the same in 2 or 3 dimensions.
Some other interesting perspectives are the presence of traps and the effects of temperature.
These are currently under investigation and are postponed for future publications.

Finally, as all these effects occur in very narrow windows of energy and density, their experimental checks are quite challenging. 
One may imagine special setups to finely adapt the measurements to such situations. In particular, the comparison of the excitation 
spectra reveals that one may select the phonon region where the frequencies are well separated. 

M. B. and A. H. wish to thank G. Shlyapnikov for fruitful discussions on the subject of correlations in BEC.

This work has been supported by The Faculty of Exact Sciences and Informatics and the Laboratory for Theoretical Physics and 
Material Physics, Hassiba Benbouali University of Chlef, Ministry of Higher Education and Scientific Research, through the 
project grants B00L02UN020120180002.


\begin{thebibliography}{99}

\bibitem{cabrera18} Cabrera C R, Tanzi L, Sanz J, Naylor B, Thomas P, Cheiney P and Tarruell L, 2018 Science 359 301.

\bibitem{semeghini18} Semeghini G, Ferioli G, Masi L, Mazzinghi C, Wolswijk L, Minardi F, Modugno M, Modugno G, Inguscio M and Fattori M, 
2018 Phys. Rev. Lett. 120 235301.

\bibitem{lee18} Lee K L et al 2018 New J. Phys. 20 053004.

\bibitem{LHY57} Lee T D, Huang K, and Yang C N, 1957 Phys. Rev. 106, 1135.

\bibitem{Petrov15} Petrov D S, 2015 Phys. Rev. Lett. 115 155302. 

\bibitem{dropletexp} D’Errico C, Burchianti A, Prevedelli M, Salasnich L, Ancilotto F,
Modugno M, Minardi F, and Fort C, 2019 Phys. Rev. Res. 1, 033155.

\bibitem{Griffin96} Griffin A, 1996 Phys. Rev. B 53, 9341.

\bibitem{olsen02} Olsen M K, Plimak L I, Kruglov V I, Hope J J, Drummond P D, and Collett M J, 2002 Laser Physics, Vol. 12, No. 1,  pp. 21–36.

\bibitem{buljan05} Buljan H, Segev M and Vardi A, 2005 Phys. Rev. Lett. 95, 180401.

\bibitem{yuk07} Yukalov V I, 2007 J. Phys. Stud. 11 55-62.

\bibitem{merh06} Merhasin I M, Malomed B A and Band Y B, 2006 Phys. Rev. A 74, 033614.

\bibitem{axel15} Lode A U J , Chakrabarti B and Kota V K B, 2015 Phys. Rev. A 92, 033622.

\bibitem{HUTCH} Hutchinson D A W, Dodd R J and Burnett K, 1998 Phys. Rev. Lett. 81 (11).


\bibitem{Dalfovo99} Dalfovo F, Giorgini S, Pitaevskii L P and Stringari S, 1999 Rev. Mod. Phys. 71, 463.

\bibitem{KB11} Kouidri S and Benarous M, 2011 J. Phys. B44, 205301.

\bibitem{BM13} Benarous M, 2013 Eur. Phys. J. D67, 243.

\bibitem{BB11} Boudjemaa A and Benarous M, 2011 Phys. Rev. A84, 043633.

\bibitem{RPGA20} Roy A, Pal S, Gautam S, Angom D and Muruganandam P, 2020 Comp. Phys. Com. Volume 256, 107288.

\bibitem{esry97} Esry B D, Greene C H, Burke J P and Bohn J L, 1997 Phys. Rev. Lett. 78, 3594.

\bibitem{tomas03} Tommasini P, de Passos E J V, de Toledo Piza A F R and Hussein M S, 2003 Phys. Rev. A67, 023606.

\bibitem{alon07} Ofir E Alon, Alexej I Streltsov and L S Cederbaum, 2007 Phys. Rev. A76, 062501.

\bibitem{bert13} Van Schaeybroeck B, 2013 Physica A392  3806–3811.

\bibitem{roy14} Roy A, Gautam S and Angom D, 2014 Phys. Rev. A89, 013617.

\bibitem{roy15} Roy A and Angom D, 2015 Phys. Rev. A92, 011601(R).

\bibitem{boudjemaa18} Boudjemaa A, 2018 Phys. Rev. A97, 033627.

\bibitem{BL58} Beliaev S T, 1958 Soviet. Phys. JETP 7, 289.

\bibitem{bogol04} Bogoliubov N M, Bullough R K, Malyshev C and Timonen J, 2004 Phys.Rev. A69, 023619.

\bibitem{bidasyuk18} Bidasyuk Y M, Weyrauch M , Momme M and Prikhodko O O, 2018 J. Phys. B: At. Mol. Opt. Phys. 51, 205301 (9pp).

\bibitem{cikojevi18} Cikojevi V et al 2018 New J. Phys. 20 085002.

\bibitem{BPfau19} B$\ddot{\rm o}$ttcher F, Wenzel M, Schmidt J N, Mingyang Guo, Langen T, Ferrier-Barbut T, and Pfau T, 2019 Phys. Rev. Res. 1, 033088.  	

\bibitem{hodgman11} Hodgman S S et al., 2011 Science 331, 1046–1049.

\bibitem{haller11} Haller E, Rabie M, Mark M J, Danz J G, Hart R, Lauber K, Pupillo G and Nagerl H C, 2011 Phys. Rev. Lett. 107, 230404.

\bibitem{dall3} Dall R G et al., 2013 Nat. Phys. 9, 341–344.

\bibitem{endres13} Endres M et al., 2013 Appl. Phys. B 113, 27–39.

\bibitem{langen15} Langen T et al., 2015 Science, 348(6231):207-2011.

\bibitem{eto16} Eto Y, Takahashi M, Nabeta K, Okada R, Kunimi M, Saito H and Hirano T, 2016 Phys. Rev. A93, 033615.

\bibitem{schweigler17} Schweigler T et al., 2017 Nature, 545(7654):323–326.

\bibitem{yang18} Yang T L, Grišins P, Chang Y T, Zhao Z H, Shih C Y, Giamarchi T and Hulet R G, 2018 Phys. Rev. Lett. 121, 103001.

\bibitem{sch19} Schweigler T, "Correlations and dynamics of tunnel-coupled one-dimensional Bose gases". PhD thesis, TU Wien, March 2019, 169 pp.

\bibitem{holland01} Holland M, Park J, and Walser R, 2001 Phys. Rev. Lett. 86, 1915.

\bibitem{cherny06} Cherny A Y and Brand J, 2006 Phys. Rev. A 73, 023612.

\bibitem{cc05} Calabrese P and Cardy J, 2005 J. of Stat. Mech.: Theory and Experiment, P04010.

\bibitem{eckart08} Eckart M, Walser R and Schleich W P, 2008 New J. Phys. 10, 045024 (28pp).

\bibitem{entang09} Horodecki R, Horodecki P, Horodecki M and Horodecki K, 2009 Rev. Mod. Phys. 81, 865.

\bibitem{mathey09} Mathey L, Danshita I and Clark C W, 2009 Phys. Rev. A79, 011602(R).

\bibitem {guan13} Guan X W, Batchelor M T and Lee C, 2013 Rev. Mod. Phys. 85, 1633.

\bibitem{johannes14} Hofmann J, Natu S S and Sarma S D, 2014 Phys. Rev. Lett. 113, 095702.

\bibitem{natu13} Natu S S, Sarma S D, 2014 Phys. Rev. Lett. 113 095702.

\bibitem{kla15} Klaiman S and Alon O E, 2015 Phys. Rev. A91, 063613.

\bibitem{liu15} Liu Boyang and Hu Jiangping, 2015 Phys. Rev. A92, 013606.

\bibitem{nandani16} Nandani EJKP, Romer R A, Tan S and Guan Xi-Wen, 2016 New J. Phys. 18, 055014.

\bibitem{bisset18} Bisset R N, Kevrekidis P G and Ticknor C, 2018 Phys. Rev. A97, 023602.

\bibitem{xi18} Xi Kui-Tian, Byrnes T and Saito H, 2018 Phys. Rev. A97, 023625.

\bibitem{ng19} King Lun Ng, Opanchuk B, Reid M D and Drummond P D, 2019 Phys. Rev. Lett. 122, 203604.

\bibitem{WB03} Whitlock N K and Bouchoule I, 2003 Phys. Rev. A68, 053609.

\bibitem{OG20} Ota M and Giorgini S, 2020 Phys. Rev. A102, 063303.

\bibitem{penna18} Penna V and Richaud A, 2018 Nature Scientific Reports 8, 10242.

\bibitem{wang16} Wang W, Penna V and B Capogrosso-Sansone, 2016 New J. Phys. 18, 063002.

\bibitem{lingua18} Lingua F, Richaud A and Penna V, 2018 Entropy, 20, 84.

\bibitem{links18} Links J and Shen Y, 2018 J. Phys. B: At. Mol. Opt. Phys. 51, 095302 (8pp).

\bibitem{OA20} Ota M and Astrakharchik G E, 2020 SciPost Phys. 9, 020.

\bibitem{zin221} Zin P and Pylak M, 2022 New J. Phys., 24, 113038.

\bibitem{zin222} Zin P, Pylak M and Gajda M, 2022 Phys. Rev. A106, 013320.

\bibitem{Sarkar20} Sarkar S, McEndoo S, Schneble D and Daley A J, 2020 New J. Phys. 22, 083017

\bibitem{alon17} Alon O E, 2017 J. Phys. A: Math. Theor. 50, 295002 (24pp).

\bibitem{bouvrie14} Bouvrie P A, Majtey A P, Tichy M C, 2014 Eur. Phys. J. D68: 346. 

\bibitem{kla17} Klaiman S, Streltsov A I and Alon O E, 2017 Chem. Phys. 482, 362.

\bibitem{kro13} Kronke S, Cao L, Vendrell O and Schmelcher P, 2013 New J. of Phys. 15, 063018 (17pp).

\bibitem{cao13} Cao L, Kronke S, Vendrell O and Schmelcher P,2013 J. Chem. Phys. 139, 134103.

\bibitem{cao17} Cao L, Bolsinger V, Mistakidis S I, Koutentakis G M, Kronke S, Schurer J M, and Schmelcher P, 2017 J. 
Chem. Phys. 147, 044106.

\bibitem{mistakidis18} Mistakidis S I, Katsimiga G C, Kevrekidis P G and Schmelcher P, 2018 New J. Phys. 20, 043052.

\bibitem{pfl09} Pflanzer A C, Zöllner S and Schmelcher P, 2009 J. Phys. B: At. Mol. Opt. Phys. 42, 231002.

\bibitem{CGW62} Tulub A V , 1961 Zh. Eksp. Teor. Fiz. 41, 1828 [Sov. Phys. JETP 14, 1301 (1962)].

\bibitem{CGW93} Altanhan T and Kandemir B S, 1993 J. Phys.: Condens. Matter 5, 6729.

\bibitem{CGW94} Kandemir B S and Altanhan T, 1994 J. Phys.: Condens. Matter 6, 4505.

\bibitem{CGW16} Shchadilova Y E, Grusdt F, Rubtsov A N and Demler E, 2016 Phys.Rev. A93, 043606.

\bibitem{BV} Balian R and V\'en\'eroni M, 1988 Ann. of Phys. (N.Y.) 187, 29.

\bibitem{BM91} Benarous M, thesis IPN-Orsay-France, October 1991.

\bibitem{BF99} Benarous M and Flocard H, 1999 Ann. of Phys. (N.Y.) 273, 242.

\bibitem{PetAst16} Petrov D S and Astrakharchik G E, 2016 Phys. Rev. Lett. 117 100401.

\bibitem{LB20} Lavoine L and Bourdel T, 2020 Phys. Rev. Lett. 117 100401.

\bibitem{TGAM20} Tilutki M, Astrakharchik G E, Malomed B A and Petrov D S, 2020 Phys. Rev. A101, 051601(R).

\bibitem{SBSS23} Sinha Sudip, Biswas S, Santos L and Sinha S, 2023 ArXiv. /abs/2304.04261.

\bibitem{EMS23} Englezos I A, Mistakidis S I and Schmelcher P, 2023 Phys. Rev. A107, 023320.

\bibitem{MOR} Morgan S A, 2000 J. Phys. B33, 3847.

\bibitem{CHER03} Chernyak V, Choi S and Mukamel S, 2003 Phys. Rev. A67, 053604.


\bibitem{min74} Mineev V P, 1974 Zh. Eksp. Teor. Fiz. 67, 263-272.




\disregard{
\bibitem{GP} E. P. Gross, Nuovo Cimento {\bf 20} (1961), 454; L. Pitaevskii,
Soviet Phys. JETP {\bf 13} (1961), 451.
\bibitem{HU87} K. Huang, {\it Statistical Mechanics} (New York: Wiley), 1987; 
A. Bulgac and Y. Yu,  Phys. Rev. Lett. 88 042504 (2002).
\bibitem{Pric} L Pricoupenko, Phys. Rev. {\bf A84} (2011), 053602; L Pricoupenko, Phys. Rev. {\bf A70} (2004), 013601; M
Olshanii and L Pricoupenko, Phys. Rev. Lett {\bf 88}, 010402 (2002).
\bibitem{KY06} T. Kita, J. Phys. Soc. Jpn {\bf 75}, 044603 (2006); V. I. Yukalov and H. Kleinert, Phys. Rev. {\bf A75}, 063612 (2006).
\bibitem{JIN97} D. S. Jin, M. R. Matthews, J. R. Ensher, C. E. Wieman and E. A. Cornell, Phys. Rev. Lett. {\bf 78}, 764 (1997).
\bibitem{TT97} E. Timmermans, P. Tommasini and K. Huang, Phys. Rev. {\bf A55} (1997), 3645.
\bibitem{RS80} P. Ring and P. Schuck, The Nuclear Many-Body Problem (Springer-Verlag, New York 1980).
\bibitem{SV00} P. Schuck and X. Vinas, Phys. Rev. {\bf A61} (2000), 43603
\bibitem{PROUK96} N. P. Proukakis and K. Burnett, J. Res. Natl. Stand. Technol. {\bf 101}, 457 (1996). 
\bibitem{YUK} V I Yukalov, Phys. Rev. {\bf E72}, 066119 (2005); Ann. of Phys.(NY) {\bf 323}, 461 (2008);
Phys. Lett. A 359, {\bf 712} (2006); Laser Phys. Lett. {\bf 3}, 406 (2006); V I and E P Yukalova, 
Laser Phys. Lett. {\bf 2}, 506 (2005).
\bibitem{KOH02} T. Kohler and K. Burnett, Phys. Rev. {\bf A65}, 033601 (2002).
\bibitem{GI04} C. Gies, M. D. Lee and D. A. W. Hutchinson, J. Phys. B: At. Mol. Opt. Phys. {\bf 38}, 1797 (2005).
\bibitem{BCB99} G. Bruun, Y. Castin and K. Burnett, Eur. Phys. J. {D7} 433 (1999).
\bibitem{HP} N.M. Hugenholtz and D. Pines, Phys. Rev. {\bf 116} (1959) 489.
\bibitem{GR} A. Griffin and H. Shi, Phys. Rep. {\bf 304} (1998), 1.
\bibitem{GS03} D.M. Gangardt and G.V. Shlyapnikov, Phys. Rev. Lett. {\bf 90}, (2003) 010401; 
D.M. Gangardt and G.V. Shlyapnikov, New Journal of Physics {\bf 5} (2003) 79. 
\bibitem{NE} J. Kneur, A. Neveu and M. B. Pinto, Phys. Rev. {\bf A69}, 053624 (2004)
\bibitem{BM98} M. Benarous, Ann. of Phys. (N.Y.) {\bf 269} (1998), 107.
\bibitem{BM98P} M Benarous, Ann. of Phys. (N.Y.) {\bf 264} (1998), 1.
\bibitem{BM05} M Benarous, Ann. of Phys. (N.Y.) {\bf 320} (2005), 226.
\bibitem{BCH08} M Benarous and H Chachou-Samet, Eur. Phys. J. D 50, 125–132 (2008).
\bibitem{BB10} A Boudjemaa and M Benarous, Eur. Phys. J. D 59, 427–434 (2010).
\bibitem{and95} M. H. Anderson, J. R. Ensher, M. R. Matthews, C. E. Wieman
and E. A. Cornell, Science {\bf 269}, 198 (1995).
\bibitem{dav95} K. B. Davis, M. O. Mewes, M. R. Andrews, N. J. Van Druten,
D. S. Durfee, D. M. Kurn and W. Ketterle, Phys. Rev. Lett. {\bf 75}, 3969
(1995).
\bibitem{BE24} S. N. Bose, Z. Phys. {\bf 26}, 178 (1924); A. Einstein,Akad.
Wiss. 1924, 261.
\bibitem{KR} W. Krauth, Phys. Rev. Lett. {\bf 77}, (1996), 3695; A. B.
Kuklov N. Chencinski, A. M. Levine, W. M. Schreiber and J. L. Birman, Phys.
Rev. {\bf A55}, (1997), 488; N. V. Prokof'ev, B. V. Svistunov and I. S.
Tupitsyn, Sov. Phys. -{\bf JETP87}, 310 (1998).
\bibitem{BG} N. Bogoliubov, J. Phys. USSR {\bf 11} (1947), 23; A. L. Fetter
and J. D. Walecka, ''Quantum Theory of Many-Particle Systems'', McGraw-Hill,
NY, 1971; L. Pitaevskii and S. Stringari, ''Bose-Einstein Condensation'',
International Series of Monographs on Physics, Oxford Science Publications,
Clarendon Press, Oxford, 2003; C. J. Pethick and H. Smith, ''Bose-Einstein
Condensation in Dilute Gases'', Cambridge University Press, Cambridge, UK,
2002.
\bibitem{PO} V. N. Popov, Sov. Phys. JETP {\bf 20} (1965), 1185;
''Functional Integrals and Collective Excitations'', Cambridge Univ. Press,
Cambridge, 1987; D. A. W. Hutchinson and E. Zaremba, Phys. Rev. {\bf A57}
(1998), 1280; D. A. W. Hutchinson E. Zaremba and A. Griffin, Phys. Rev.
Lett. {\bf 78} (1997), 1842; S. A. Gardiner, Phys. Rev. {\bf A56} (1997),
1414.
\bibitem{CAR09} M. A. Caracanhas, J. A. Seman, E. R. Ramos, E. A. L. Henn, K. M. F. Magalhaes, K. Helmerson and V. S. Bagnato, J. Phys. {\bf B42} 145304 (2009).
\bibitem{PG97} V. M.Perez-Garcia, H. Michinel, J. I. Cirac, M. Lewenstein and P. Zoller, Phys.
Rev. {\bf A56}, (1997), 1424; Phys. Rev. Lett. {\bf 77} (1996), 5320.
\bibitem{KT97} A. K. Kerman and P. Tommasini, Phys. Rev. {\bf B56}, (1997),
14733; Ann. of Phys. (N.Y.) {\bf 260} (1997), 250.
\bibitem{S98} S. Stenholm, Phys. Rev. {\bf A57} (1998), 2942.
\bibitem{burn2004} J. Rogel-Salazar, S. Choi, G. H. C. New and K. Burnett, J. Opt. B: Quantum Semiclass. Opt. 6 
(2004) R33–R59. 
\bibitem{dehk17} A S Dehkharghani, F F Bellotti and N T Zinner, J. Phys. B: At. Mol. Opt. Phys. 50 (2017) 144002 (11pp).
\bibitem{prz18} Przemysaw Kocik et al 2018 EPL 123 36001.
\bibitem{Hutchinson97} D A W Hutchinson, E Zaremba and A Griffin, Phys. Rev. Lett. 78, 1842 (1997). 
\bibitem{boudjemaa15} A Boudjemaa, J. Phys. A: Math. Theor. 48 (2015) 045002.
\bibitem{guebli19} N Guebli and A Boudjemaa, J. Phys. B: At. Mol. Opt. Phys. 52 (2019). 
\bibitem{yuka14} V I Yukalov and E P Yukalova, J. Phys. B: At. Mol. Opt. Phys. 47 (2014) 095302 (6pp).
\bibitem{yuk16} V I Yukalov, Laser Phys. 26 (2016) 062001 (74pp).
\bibitem{yuka16} V I Yukalov and E P Yukalova, Laser Phys. 26 (2016) 045501 (14pp).

}


\end{thebibliography}
\end{document}